\let\accentvec\vec
\documentclass[a4paper]{llncs}

\let\vec\accentvec
\usepackage{amsmath,amssymb,wasysym,bbm,url,ifthen,calc,hyperref}

\let\phi\varphi
\let\epsilon\varepsilon
\let\theta\vartheta
\let\mathbb\mathbbm

\newcommand{\NN}{\mathbb{N}}

\newcommand{\RR}{\mathbb{R}}
\newcommand{\CC}{\mathbb{C}}

\renewcommand{\ker}{\operatorname{Ker}}
\newcommand{\im}{\operatorname{Im}}
\newcommand{\orth}[1]{#1^\perp}
\newcommand{\biorth}[1]{#1^{\perp\perp}}

\newcommand{\diag}{\operatorname{diag}}
\newcommand{\mat}[3]{#1^{#2}_{#3}}

\newcommand{\finset}{\smash{\NN^\circledast}}
\newcommand{\pow}{\mathcal{P}}
\newcommand{\Ev}{\operatorname{Ev}}

\newcommand{\der}{D}
\newcommand{\cum}{{\textstyle \varint}}
\newcommand{\fri}[1]{#1^\Diamond}
\newcommand{\trc}{\operatorname{trc}}

\newcommand{\galg}{\mathcal{F}}
\newcommand{\ogalg}{\mathcal{G}}
\newcommand{\oogalg}{\mathcal{H}}
\newcommand{\func}[2]{\galg^{#1}_{#2}}
\newcommand{\bspc}{\mathcal{B}}
\newcommand{\bvp}[2]{\boxed{\begin{array}{l}#1\\#2\end{array}}}

\spnewtheorem{myexample}[theorem]{Example}{\bfseries}{\upshape}
\spnewtheorem{mydefinition}[theorem]{Definition}{\bfseries}{\upshape}
\spnewtheorem{mylemma}[theorem]{Lemma}{\bfseries}{\itshape}
\spnewtheorem{myproposition}[theorem]{Proposition}{\bfseries}{\itshape}
\newenvironment{myproof}{\begin{proof}}{\qed\end{proof}}

\title{A Symbolic Approach to Boundary Problems for Linear Partial Differential
  Equations} \subtitle{Applications to the Completely Reducible Case of the
  Cauchy Problem with Constant Coefficients} \titlerunning{Symbolic Theory of
  Boundary Problems with Applications} \author{Markus Rosenkranz \and Nalina
  Phisanbut\thanks{The authors acknowledge support from the EPSRC First Grant
    EP/I037474/1.}\\
  \small\email{\{M.Rosenkranz,N.Phisanbut\}@kent.ac.uk}} \institute{University
  of Kent, Canterbury, Kent CT2 7NF, United Kingdom}

\begin{document}
\maketitle
\thispagestyle{plain}
\pagestyle{plain}

\begin{abstract}
  We introduce a general algebraic setting for describing linear boundary
  problems in a symbolic computation context, with emphasis on the case of
  partial differential equations. The general setting is then applied to the
  Cauchy problem for completely reducible partial differential equations with
  constant coefficients. While we concentrate on the theoretical features in
  this paper, the underlying operator ring is implemented and provides a
  sufficient basis for all methods presented here.
\end{abstract}

\section{Introduction}
\label{sec:intro}

A symbolic framework for boundary problems was built up
in~\cite{Rosenkranz2005,RosenkranzRegensburger2008a} for linear ordinary
differential equations (LODEs); see
also~\cite{RosenkranzRegensburgerTecBuchberger2012,KorporalRegensburgerRosenkranz2011,KorporalRegensburgerRosenkranz2012}
for more recent developments. One of our long-term goals is to extend this to
boundary problems for linear partial differential equations (LPDEs). Since this
is a daunting task in full generality, we want to tackle it in stages of
increasing generality. In the first instance, we restrict ourselves to
\emph{constant coefficients}, where the theory is quite
well-developed~\cite{Hormander1976}. Within this class we distinguish the
following three stages:
\begin{enumerate}
\item The simplest is the Cauchy problem for \emph{completely reducible}
  operators.
\item The next stage will be the \emph{Cauchy problem} for general hyperbolic
LPDEs.
\item After that we plan to study \emph{boundary problems} for
  elliptic/parabolic LPDEs.
\end{enumerate}

In this paper we treat the first case (Section~\ref{sec:cauchy-analytic}). But
before that we build up a \emph{general algebraic framework}
(Sections~\ref{sec:boundary-data} and~\ref{sec:greens-operators}) that allows a
symbolic description for all boundary problems (LPDEs/LODEs, scalar/system,
homogeneous/inhomogeneous, elliptic/hyperbolic/parabolic). Using these concepts
and tools we develop a general solution strategy for the Cauchy problem in the
case~(1). See the Conclusion for some thoughts about the next two steps.

\noindent The passage from LODEs to LPDEs was addressed at two earlier occasions:
\begin{itemize}
\item An \emph{abstract theory of boundary problems} was developed
  in~\cite{RegensburgerRosenkranz2009}, including LODEs and LPDEs as well as
  linear systems of these. The concepts and results of
  Sections~\ref{sec:boundary-data} and~\ref{sec:greens-operators} are built on
  this foundation, adding crucial concepts whose full scope is only appreciated
  in the LPDE setting: boundary data, semi-homogeneous problem, state operator.
\item An algebraic language for \emph{multivariate differential and integral
    operators} was introduced in Section~4
  of~\cite{RosenkranzRegensburgerTecBuchberger2009}, with a prototype
  implementation described in Section~5 of the same paper. This language is
  generalized in the PIDOS algebra of Section~\ref{sec:cauchy-analytic}, and it
  is also implemented in a Mathematica package.
\end{itemize}

In this paper we will not describe the current state of the
\emph{implementation} (mainly because of space limitations). Let us thus say a
few words about this here. A complete reimplementation of the PIDOS package
described in~\cite{RosenkranzRegensburgerTecBuchberger2009} is under way. The
new package is called OPIDO (Ordinary and Partial Integro-Differential
Operators), and it is implemented as a standalone Mathematica package unlike its
predecessor, which was incorporated into the THEOREMA system. In fact, our
reimplementation reflects several important design principles of THEOREMA,
emphasizing the use of functors and a strong support for modern two-dimensional
(user-controllable) parsing rules. We have called this programming paradigm
FUNPRO, first presented at the Mathematica symposium~\cite{Rosenkranz2012}. The
last current stable version of the (prototype) package can be found
at~\url{http://www.kent.ac.uk/smsas/personal/mgr/index.html}.

At the time of writing, the ring of \emph{ordinary integro-differential
  operators} is completed and the ring of \emph{partial integro-differential
  operators} is close to completion (for two independent variables). Compared
to~\cite{RosenkranzRegensburgerTecBuchberger2009}, the new PIDOS ring contains
several crucial new rewrite rules (instances of the substitution rule for
resolving multiple integrals). Our conjecture is that the new rewrite system is
noetherian and confluent but this issue will be analyzed at another occasion.

\emph{Notation}. The algebra of~$m \times n$ matrices over a field~$K$ is
written as~$\mat{K}{m}{n}$, where~$m=1$ or~$n=1$ is omitted. Thus we
identify~$K^m = K \oplus \cdots \oplus K$ with the space of column vectors
and~$K_n = (K^n)^*$ with the space of row vectors. More generally, we
have~$\mat{K}{m}{n} \cong K^n \to K^m$.

\section{An Algebraic Language for Boundary Data}
\label{sec:boundary-data}

As mentioned in the Introduction, we follow the \emph{abstract setting}
developed in~\cite{RegensburgerRosenkranz2009}. We will motivate and
recapitulate some key concepts here, but for a fuller treatment of these issues
we must refer the reader to~\cite{RegensburgerRosenkranz2009} and its references.

Let us recall the notion of boundary problem. Fix vector
spaces~$\galg$ and~$\ogalg$ over a common ground field~$K$
of characteristic zero (for avoiding trivialities one may
assume~$\galg$ and~$\ogalg$ to be
infinite-dimensional). Then a \emph{boundary problem}~$(T, \bspc)$
consists of an epimorphism~$T\colon \galg \to \ogalg$ and a
subspace~$\bspc \subseteq \galg*$ that is orthogonally closed in
the sense defined below. We call~$T$ the differential operator
and~$\mathcal{B}$ the \emph{boundary space}.

Similar to the correspondence of ideals/varieties in algebraic geometry, we make
use of the following Galois
connection~\cite[A.11]{RegensburgerRosenkranz2009}. If~$\mathcal{A}$ is any
subspace of the space~$\galg$, its \emph{orthogonal} $\orth{\mathcal{A}} \le
\galg^*$ is defined as~$\{ \phi \in \galg^* \mid \phi(a) = 0 \text{ for all $a
  \in \mathcal{A}$} \}$. Dually, for a subspace~$\mathcal{B}$ of the dual
space~$\galg^*$, the orthogonal~$\orth{\mathcal{B}} \le \galg$ is defined by~$\{
f \in \galg \mid \beta(f) = 0 \text{ for all $\beta \in \mathcal{B}$} \}$. If we
think of~$\galg$ as ``functions'' and of~$\galg^*$ as ``boundary conditions'',
then~$\orth{\mathcal{A}}$ is the space of valid conditions (the boundary
conditions satisfied by the given functions) while~$\orth{\mathcal{B}}$ is the
space of admissible functions (the functions satisfying the given conditions).

Naturally, a subspace of either~$\mathcal{S}$ of~$\galg$ or~$\galg^*$ is called
\emph{orthogonally closed} if $\biorth{\mathcal{S}} = \mathcal{S}$. But while
any subspace of~$\galg$ itself is always orthogonally closed, this is far from
being the case of the subspaces of the dual~$\galg^*$. Hence the condition on
boundary spaces~$\mathcal{B}$ to be orthogonally closed is in general not
trivial. However, if~$\mathcal{B}$ is finite-dimensional as in boundary problems
for LODEs (as in Example~\ref{ex:heat-conduction} below), then it is
automatically orthogonally closed. For LPDEs, the condition of orthogonal
closure is important; see Example~\ref{ex:wave-inhom-medium} for an intuitive
explanation.

In~\cite{Rosenkranz2005,RosenkranzRegensburger2008a} and also in the abstract
setting of~\cite{RegensburgerRosenkranz2009} we have only considered what is
sometimes called the \emph{semi-inhomgeneous boundary
  problem}~\cite{Stakgold1979}, more precisely the semi-inhomogeneous
incarnation of~$(T, \mathcal{B})$; see Definition~\ref{def:transfer-operator}
for the full picture. This means we are given a \emph{forcing function} $f \in
\ogalg$ and we search for a solution~$u \in \galg$ with
\begin{equation}
  \bvp{Tu=f,}{\beta(u) = 0 \; (\beta \in \mathcal{B}).}
\end{equation}
In other words, $u$ satisfies the inhomogeneous ``differential equation'' $Tu =
f$ and the homogeneous ``boundary conditions'' $\beta(u) = 0$ given
in~$\mathcal{B}$.

A boundary problem which admits a unique solution~$u \in \galg$ for every
forcing function~$f \in \ogalg$ is called \emph{regular}. In terms of the
spaces, this condition can be expressed equivalently by requiring that~$\ker{T}
\dotplus \orth{\bspc} = \galg$; see~\cite{RegensburgerRosenkranz2009} for
further details. In this paper we shall deal exclusively with regular boundary
problems. For singular boundary problems we refer the reader
to~\cite{KorporalRegensburgerRosenkranz2011} and~\cite{Korporal2012}.

For a regular boundary problem, one has a linear operator~$G\colon \ogalg \to
\galg$ sending~$f$ to~$u$ is known as the \emph{Green's operator} of the
boundary problem~$(T, \mathcal{B})$. From the above we see that~$G$ is
characterized by~$TG=1$ and~$\im{G} = \orth{\mathcal{B}}$.

\begin{myexample}
  \label{ex:heat-conduction}
  A classical example of this notion is the two-point boundary problem. As a
  typical case, consider the simplified model of \emph{stationary heat
    conduction} described by
  \begin{equation*}
    \bvp{u''=f,}{u(0) = u(1) = 0.}
  \end{equation*}
  Here we can choose~$\galg = \ogalg = C^\infty(\RR)$ for the function space
  such that the differential operator is given by $T = D^2\colon C^\infty(\RR)
  \to C^\infty(\RR)$ and the boundary space by the two-dimensional subspace
  of~$C^\infty(\RR)^*$ spanned by the linear functionals~$L\colon u \mapsto
  u(0)$ and~$R\colon u \mapsto u(1)$ for evaluation on the left and right
  endpoint. In the sequel we shall write~$\mathcal{B} = [L, R]$, employing an
  important generalization for LPDEs, described in the next eamples. We can
  express its Green's operator in the language of integro-differential operators
  as explained in~\cite{RosenkranzRegensburger2008a}.
\end{myexample}

\begin{myexample}
  \label{ex:wave-inhom-medium}
  As a typical counterpart in the world of LPDEs, consider the
  \emph{equation for waves in an inhomogeneous medium}, described by
  \begin{equation*}
    \bvp{u_{xx}-u_{tt} = f(x,t)}{u(x,0) = u_t(x,0) = u(0,t) = u(1,t) = 0}
  \end{equation*}
  in one space dimension. In this case we choose~$\galg = \ogalg = C^\omega(\RR
  \times \RR^+)$; again one could choose much larger spaces of functions (or
  distributions) in analysis and in the applications. Here the differential
  operator is~$T = \der_{xx}-\der_{tt}\colon C^\omega(\RR \times \RR^+) \to
  C^\omega(\RR \times \RR^+)$ while the boundary space~$\bspc$ is the orthogonal
  closure of the linear span of the families of functionals~$\beta_x, \gamma_x
  \; (x \in \RR)$ and~$\kappa_t, \lambda_t ; (t \in \RR^+)$ defined
  by~$\beta_x(u) = u(x,0), \gamma_x(u) = u_t(x,0)$ and~$\kappa_t(u) = u(0,t),
  \lambda_t(u) = u(1,t)$. Using the notation $[\dots]$ for denoting the
  orthogonal closure of the linear span, we can thus write~$\mathcal{B} =
  [\beta_x, \gamma_x, \kappa_t, \lambda_t \mid x \in \RR, t \in \RR^+]$ for the
  boundary space under consideration.

  The point of the \emph{orthogonal closure} is that the given conditions imply
  other conditions not in their span, for example~$u_x(1/2,0) = 0$
  or~$\smash{\cum_{-3}^{5}} u(0,\tau) \, d\tau = 0$. Rather than being linear
  consequences, these two examples are differential and integral
  consequences. (Of course the full boundary space also contains many
  functionals without a natural analytic interpretation.)
\end{myexample}

In the problems above, the differential equation is inhomogeneous while the
boundary conditions are homogeneous. A \emph{semi-homogeneous boundary problem}
is the opposite, combining a homogeneous differential equation with
inhomogeneous boundary conditions. While this is a simple task for LODEs (as
always we assume that the fundamental system is available to us in some form!),
it is usually a nontrivial problem for LPDEs (even when they have constant
coefficients). We will give the formal definition of a semi-inhomogeneous
boundary problem in the next section
(Definition~\ref{def:transfer-operator}). Here it suffices to consider an
example for developing the necessary auxiliary notions.

\begin{myexample}
  \label{ex:cauchy-wave-eq}
  The \emph{Cauchy problem for the wave equation} in one dimension is
  \begin{equation*}
    \bvp{u_{xx} - u_{tt} = 0,}{u(x,0) = f(x), u_t(x,0) = g(x).}
  \end{equation*}
  Being a hyperbolic problem, we could use rather general function spaces for
  the ``boundary data'' $f,g$. For reasons of uniformity we will nevertheless
  restrict ourselves here to the analytic setting, so assume~$f,g \in
  C^\omega(\RR)$. Note that the association of~$u$ to~$(f,g)$ is again a linear
  operator mapping (two univariate) functions to a (bivariate) function; we will
  come back to this point in Definition~\ref{def:transfer-operator}.
\end{myexample}

Going back to the abstract setting, one is tempted to define the notion of
boundary data as some kind of functions depending on ``fewer'' variables. But
the problem with this approach is that---abstractly speaking---we are not
dealing with any functions depending on any number of variables (but see
below). Moreover, the inhomogeneous boundary conditions in the
form~$u(0,x)=f(x), u_t(0,x)=g(x)$ are basis-dependent while the whole point of
the abstract theory is to provide a \emph{basis-free description} (which leads
to an elegant setting for describing composition and factorization of abstract
boundary problems); see Proposition~\ref{prop:bp-product}. We shall therefore
develop a basis-independent notion of boundary data (we can go back to the
traditional description by choosing a basis).

We define first the \emph{trace map}~$\trc\colon \galg \to \bspc^*$ as
sending~$f \in \galg$ to the functional~$\beta \mapsto \beta(f)$. In
Example~\ref{ex:cauchy-wave-eq} this would map the function~$u(x,t)$ to its
position and velocity values on~$\RR \times \{0\}$. We call~$\trc(f)$ the trace
of~$f$ and write it as~$f^*$. Moreover, we denote the image of the map~$\trc$
by~$\bspc'$ and refer to its elements as \emph{boundary data}. Note
that~$\bspc'$ is usually much smaller than the full dual~$\bspc^*$ since a
continuous function (let alone an analytic one) cannot assume arbitrary
values. (This situation is vaguely reminiscent of the algebraic and continuous
dual of a topological vector space.)

Since by definition the trace map is surjective from~$\galg$ to~$\bspc'$, it has
some right inverse~$\fri{\bspc}\colon \bspc' \to \galg$. We refer
to~$\fri{\bspc}$ as an \emph{interpolator} for~$\bspc$ since it constructs a
``function''~$f = \fri{\bspc}(B) \in \galg$ from given boundary values~$B \in
\bspc'$ such that~$\beta(f) = B(\beta)$. Of course, the choice of~$f$ is usually
far from being unique. Apart from its use for describing boundary data (see at
the end of this section), the notion of interpolator will turn out to be useful
for solving the semi-homogeneous boundary problem (see
Proposition~\ref{prop:proj-transfer}).

Let us now describe how to relate these abstract notions to the usual
setting of initial and boundary values problems as they actually in
analysis: essentially by \emph{choosing a basis}. However, we have to
be a bit careful since we must deal with the orthogonal closure.

\begin{mydefinition}
  \label{def:boundary-basis}
  If~$\bspc \le \galg^*$ is any orthogonally closed subspace, we call a
  family~$(\beta_i \mid i \in I)$ a \emph{boundary basis} if~$\bspc = [\beta_i
  \mid i \in I]$, meaning~$\bspc$ is the orthogonal closure of the span of
  the~$\beta_i$.
\end{mydefinition}

Note that a boundary basis is typically smaller than a~$K$-linear
basis of~$\bspc$. All traditional boundary problems are given in terms
of such a boundary basis. In Example~\ref{ex:cauchy-wave-eq}, the
boundary basis could be spelled out by using~$I = \RR \uplus \RR$
with~$\beta_{(x,0)}(u) = u(x,0)$ and~$\beta_{(x,1)}(u) =
u_t(x,0)$. Relative to a boundary basis~$(\beta_i \mid i \in I)$, we
call~$f = \beta_i(f)_{i \in I} \in K^I$ the \emph{boundary values}
of~$f \in \galg$. As we can see from the next proposition, we
may think of the trace as a basis-free description of boundary
values. Conversely, one can always extract from any given boundary
data~$B \in \bspc'$ the boundary values~$B(\beta_i)_{i \in I}$ as its
coordinates relative to the boundary basis~$(\beta_i)$.

\begin{mylemma}
  \label{lem:trc-coord}
  Let~$\bspc \le \galg^*$ be a boundary space with boundary
  basis~$(\beta_i \mid \in I)$. If for any~$B, \tilde{B} \in \bspc'$
  one has~$B(\beta_i)_{i \in I} = \tilde{B}(\beta_i)_{i \in I}$ then
  also~$B = \tilde{B}$. In particular, for any~$f \in \galg$,
  the trace~$f^*$ depends only on the boundary values~$f(\beta_i)_{i
    \in I}$.
\end{mylemma}
\begin{myproof}
  Since~$\bspc'$ is the image under the trace map, we have~$B = f^*$
  and~$\tilde{B} = \tilde{f}^*$ for some~$f, \tilde{f} \in \galg$. So
  assume~$f(\beta_i) = \tilde{f}(\beta_i)$ for all~$i \in I$ and
  thus~$\beta(f-\tilde{f}) = 0$ for all~$\beta \in \bspc$ by the definition of
  orthogonal closure. Then we have~$\trc(f-\tilde{f}) = 0$ and thus~$f^* =
  \tilde{f}^*$.
\end{myproof}

The \emph{analytic interpretation} of this proposition is clear in
concrete cases like Example~\ref{ex:wave-inhom-medium}: Once the
values~$u(x,0), u_t(x,0)$ and~$u(0,t), u(1,t)$ are fixed, all
differential and integral consequences, as in the above
examples~$u_x(1/2,0) = 0$ or~$\cum_{1/4}^{3/4} u(0,\tau) \, d\tau$,
are likewise fixed. It is therefore natural that an interpolator need
only consider the boundary values rather than the full trace
information. This is the contents of the next lemma.

\begin{mylemma}
  \label{lem:int-coord}
  Let~$\bspc \le \galg^*$ be a boundary space with boundary
  basis~$(\beta_i \mid \in I)$ and write~$f_I = f(\beta_i)_{i \in I}
  \in K^I$ for the boundary values of any~$f \in \galg$
  and~$\bspc_I'$ for the $K$-subspace of~$K^I$ generated by all
  boundary values~$f_I$. Then any linear map~$J\colon \bspc_I' \to
  \galg$ with~$J(f_I)_I = f_I$ induces a unique
  interpolator~$\fri{\bspc}\colon \bspc' \to \galg$ defined by
  $B \mapsto J(B(\beta_i)_{i \in I})$.
\end{mylemma}
\begin{myproof}
  We must show that~$\fri{\bspc}\colon \bspc' \to \galg$ is a right inverse
  of~$\trc\colon \galg \to \bspc'$. So for arbitrary~$B = \in \bspc'$ we must
  show~$J(B(\beta_i)_{i \in I})^* = B$. By the definition of~$\bspc'$ we can
  write~$B = f^*$ for some~$f \in \galg$. Since~$f^*(\beta_i) = \beta_i(f)$, we
  are left to prove~$J(f_I)^* = f^*$. Using Lemma~\ref{lem:trc-coord}, it
  suffices to prove that~$J(f_I)_I = f_I$, which is true by hypothesis.
\end{myproof}

As noted above, we can always extract the boundary values~$B(\beta_i)_{i \in I}
\in K^I$ of some boundary data~$B \in \bspc'$ relative to fixed
basis~$(\beta_i)$ of~$\bspc$. However, since one normally has got \emph{only}
the boundary values (coming from some function), where does the corresponding~$B
\in \bspc'$ come from? By definition, it has to assign values to all~$\beta \in
\bspc$, not only to the~$\beta_i$ making up the boundary basis. As suggested by
the above lemmata, for actual computations those additional values will be
irrelevant. Nevertheless, it gives a feeling of confidence to provide these
values: If~$\fri{\bspc}$ is any \emph{interpolator}, we have~$B(\beta) =
\beta(\fri{\bspc}(B_i)_{i \in I})$. This follows immediately from the fact
that~$\fri{\bspc}$ is a right inverse of the trace map and that it depends only
on the boundary values~$(B_i)_{i \in I}$ by Lemma~\ref{lem:int-coord}. In the
analysis setting this means we interpolate the given boundary value and then do
with the resulting function whatever is desired (like derivatives and integrals
in Example~\ref{ex:wave-inhom-medium}).

\section{Green's Operators for Signals and States}
\label{sec:greens-operators}

Using the notion of boundary data developed in the previous section, we can now
give the formal definition of the \emph{semi-homogeneous boundary problem}. In
fact, we can distinguish three different incarnations of a ``boundary problem''
(as we assume regularity, the \emph{fully homogeneous problem} is of course
trivial).

\begin{mydefinition}
  \label{def:transfer-operator}
  Let~$(T, \bspc)$ be a regular boundary problem with~$T\colon \galg \to \ogalg$
  and boundary space~$\bspc \subseteq \galg^*$. Then we distinguish the
  following problems: \medskip

  \setlength{\tabcolsep}{0pt}\noindent%
  \begin{tabular}{ccc}
    \parbox{0.4\textwidth}{\small
      Given~$(f, B) \in \ogalg \oplus \bspc'$,\\
      find~$u \in \galg$ with\\[1ex]
      $\bvp{Tu = f,}{\beta(u) = B(\beta) \; (\beta \in \bspc).}$}&
    \parbox{0.27\textwidth}{\small
      Given~$f \in \ogalg$,\\
      find~$u \in \galg$ with\\[1ex]
      $\bvp{Tu = f,}{\beta(u) = 0 \; (\beta \in \bspc).}$}\kern1.1em&
    \parbox{0.29\textwidth}{\small
      Given~$B \in \bspc'$,\\
      find~$u \in \galg$ with\\[1ex]
      $\bvp{Tu = 0,}{\beta(u) = B(\beta) \; (\beta \in \bspc).}$}
  \end{tabular}
  \medskip
  
  \noindent They are, respectively, called the \emph{fully inhomogeneous}, the
  \emph{semi-inhomogeneous} and the \emph{semi-homogeneous} boundary problem
  for~$(T, \bspc)$. The corresponding linear operators will be written
  as~$F\colon \ogalg \oplus \bspc' \to \galg$, $(f,B) \mapsto u$ and~$G\colon
  \ogalg \to \galg$, $f \mapsto u$ and~$H\colon \bspc' \to \galg$, $B \mapsto
  u$.
\end{mydefinition}

\begin{mylemma}
  Each of the three problems in Definition~\ref{def:transfer-operator} has a
  unique solution for the respective input data,so the operators~$F, G, H$ are
  well-defined.
\end{mylemma}
\begin{myproof}
  Employing the usual superposition principle~$F = G \oplus H$, we can restrict
  ourselves to the semi-inhomogeneous and the semi-homogeneous boundary
  problem. For the former, the existence and uniqueness is verified
  in~\cite[\S2]{RegensburgerRosenkranz2009}. For the semi-homogeneous problem,
  existence is seen as follows: Since the given boundary data~$B \in \bspc'$ can
  be written as~$B = f^*$ by the definition of~$\bspc'$, we have to find~$u \in
  \galg$ such that~$Tu=0$ and~$\beta(u) = \beta(f)$ for all~$\beta \in
  \bspc$. Using the decomposition~$f=f_0 + f_1$ corresponding to the direct
  sum~$\ker{T} \dotplus \orth{\bspc} = \galg$, we set~$u = f_0$. Then~$Tu=0$ is
  clear, and we check further that~$\beta(u) = \beta(f)-\beta(f_1) =
  \beta(f)$. For uniqueness, it suffices to prove that~$Tu=0, u^* = 0$ has only
  the trivial solution, which is clear since the sum~$\ker{T} \dotplus
  \orth{\bspc} = \galg$ is direct.
\end{myproof}

The terminology for the operators~$F, G, H$ is not uniform in the literature. In
the past, we have only considered~$G$ and called it the ``Green's operator''
acting on a ``forcing function''~$f$. While this is in good keeping with the
engineering tradition and large parts of the standard mathematical
culture~\cite{Stakgold1979}, it is difficult to combine with suitable
terminology for~$F$ and~$H$. In this paper, we shall follow the systems theory
jargon~\cite{OberstPauer2001} and refer to~$F$ as the (full) \emph{transfer
  operator}, to~$G$ as the (zero-state) signal transfer operator or briefly
\emph{signal operator}, and to~$H$ as the (zero-signal) state transfer operator
or briefly \emph{state operator}. This terminology reflects the common view of
forcing functions~$f \in \galg$ as ``signals'' and boundary data~$B \in \bspc'$
as (initial) ``states''.

One of the advantages of the abstract formulation is that it allows us to
describe the \emph{product of boundary problems} in a succinct, basis-free
manner (and it includes LODEs and LPDEs as well as systems of these). The
composite boundary problem can then be solved, both in its semi-inhomogneous and
its semi-homogeneous incarnation (the latter is presented here for the first
time).

\begin{myproposition}
  \label{prop:bp-product}
  Define the product of two boundary problems~$(T, \bspc)$ and~$(\tilde{T},
  \tilde{\bspc})$ with~$\vphantom{\overset{e}{\to}}\smash{\galg
    \overset{\tilde{T}}{\to} \ogalg \overset{T}{\to} \oogalg}$ and~$\bspc
  \subseteq \ogalg^*$, $\tilde{\bspc} \subseteq \galg$ by
  \begin{equation*}
    (T, \bspc) (\tilde{T}, \tilde{\bspc}) = (T \tilde{T}, \bspc \tilde{T} +
    \tilde{\bspc}).
  \end{equation*}
  Then~$(T, \bspc) (\tilde{T}, \tilde{\bspc})$ is regular if both factors
  are. In that case, if~$(T, \bspc)$,~$(\tilde{T}, \tilde{\bspc})$ have,
  respectively, the signal operators~$G$,~$\tilde{G}$ and the state
  operators~$H$,~$\tilde{H}$, then~$(T, \bspc) (\tilde{T}, \tilde{\bspc})$ has
  the signal operator~$\tilde{G} G$ and the state operator~$(\bspc \tilde{T} +
  \tilde{\bspc})' \to \galg$ acting by~$B + \tilde{B} \mapsto \tilde{G}H(B
  \tilde{T}^*) + \tilde{H}(\tilde{B})$.
\end{myproposition}
\begin{myproof}
  The preservation of regularity and the relation for the signal operators is
  proved in Proposition3.2 of~\cite{RegensburgerRosenkranz2009}. For the statement
  about the composite state operator, note first that the sum~$\bspc \tilde{T} +
  \tilde{\bspc}$ is direct by~\cite[(3.2)]{RegensburgerRosenkranz2009}; hence
  the definition is consistent. Now let~$B + \tilde{B} \in (\bspc \tilde{T} +
  \tilde{\bspc})'$ be arbitrary boundary data and set~$u = \tilde{G}H(B
  \tilde{T}^*) + \tilde{H}(\tilde{B})$. Then~$T \tilde{T} u = TH(B \tilde{T}^*)
  = 0$ since~$\tilde{T}\tilde{G} = 1$ and both~$\tilde{T} \tilde{H}$ and~$TH$
  vanish by the definition of state operator. Hence the differential equation of
  the composite semi-homogeneous boundary problem is satisfied.

  It remains to check the boundary conditions~$\beta \tilde{T}(u) = B(\beta
  \tilde{T})$ for~$\beta \in \bspc$ and~$\tilde{\beta}(u) =
  \tilde{B}(\tilde{\beta})$ for~$\tilde{\beta} \in \tilde{\bspc}$. For the
  first, we use~$\tilde{T} \tilde{G} = 1$ and~$\tilde{T} \tilde{H} = 0$ again to
  compute~$\beta \tilde{T}(u) = \beta H (B \tilde{T}^*)$. Since~$\tilde{T}^*
  \kern-1.5pt B \in \bspc'$ and~$H$ is the state operator for~$(T, \bspc)$, we
  obtain~$\beta H(B \tilde{T}^*) = (B \tilde{T}^*) (\beta) = B(\beta \tilde{T})$
  as required. For the second set of boundary conditions, we use
  that~$\tilde{\beta} \tilde{G} = 0$ since~$\tilde{G}$ is the signal operator
  with homogeneous boundary conditions~$\tilde{\beta} \in
  \tilde{\bspc}$. Hence~$\tilde{\beta}(u) = \tilde{\beta} \tilde{H}(\tilde{B})$,
  and now the claim follows because~$\tilde{H}$ is the state operator for the
  inhomogeneous boundary conditions~$\tilde{\beta} \in \tilde{\bspc}$.
\end{myproof}

As detailed in~\cite{Rosenkranz2005,RegensburgerRosenkranz2009}, the computation
of the \emph{signal operator}~$G$ can be decomposed in two parts: (1) Finding a
\emph{right inverse}~$\fri{T}$ of the differential operator~$T$, which involves
only the differential equation without boundary conditions (so we may replace
the boundary by intial conditions, thus having again a unique solution: this is
the so-called fundamental right inverse). (2) Determining the \emph{projector}
onto the homogeneous solution space along the space of functions admissible for
the given boundary conditions---the projector ``twists'' the solutions coming
from the right inverse into satisfying the boundary conditions. An analogous
result holds for the computation of the \emph{state operator}~$H$ if we replace
the right inverse~$\fri{T}$ of~$T$ by the interpolator~$\fri{\bspc}$ for the
boundary space~$\bspc$.

\begin{myproposition}
  \label{prop:proj-transfer}
  Let~$(T, \bspc)$ be regular with operators~$F, G, H$ as in
  Definition~\ref{def:transfer-operator}. Then we have~$G = (1-P) \, \fri{T}$
  and~$H = P \fri{\bspc}$, hence~$F = (1-P) \, \fri{T} \oplus P \fri{\bspc}$ for
  the transfer operator.  Here~$\fri{T}\colon \ogalg \to \galg$ is any right
  inverse of the differential operator~$T\colon \galg \to \ogalg$
  and~$\fri{\bspc}\colon \bspc' \to \galg$ any interpolator for~$\bspc$
  while~$P\colon \galg \to \galg$ is the projector determined by~$\im{P} =
  \ker{T}$ and~$\ker{P} = \orth{\bspc}$.
\end{myproposition}
\begin{myproof}
  The formula for~$G$ is given in~\cite[(2.3)]{RegensburgerRosenkranz2009}. For
  proving~$H = P \fri{\bspc}$, let~$B \in \bspc'$ be arbitrary and set~$u = P
  \fri{\bspc}(B)$. Then~$Tu=0$ follows since~$\im{P} = \ker{T}$. Furthermore,
  for every~$\beta \in \bspc$ we have
  \begin{equation*}
    \beta(u) = \beta(\fri{\bspc} \! B) - \beta((1-P) \fri{\bspc} \! B) =
    B(\beta) - 0
  \end{equation*}
  by the definition of~$\fri{\bspc}$ and~$\im(1-P) = \ker{P} =
  \orth{\bspc}$. This means that~$u = H(B)$ satisfies the boundary conditions,
  so~$H$ solves the semi-homogeneous boundary problem for~$(T, \bspc)$.
\end{myproof}

If~$T$ is a completely reducible differential operator with constant
coefficients in~$\CC$, the determination of~$\fri{T}$ reduces to solving an
inhomogeneous first-order equation with constant coefficients---which is of
course straightforward (Lemma~\ref{lem:transport-equation}). Also the
determination of the interpolator~$\fri{\bspc}$ turns out to be easy for a
Cauchy problem since it is essentially given by the corresponding Taylor
polynomial~\eqref{eq:fund-intp}. Hence it remains to find some means for
computing the \emph{kernel projector}~$P$ for a boundary problem~$(T, \bspc)$.

In the case of a LODE of order~$n$, the method for computing~$P$ given in the
proof of Theorem~26 of~\cite{RosenkranzRegensburger2008a} and in Section~6
of~\cite{RegensburgerRosenkranz2009} is essentially a Gaussian elimination on
the so-called \emph{evaluation matrix}~$\beta(u) = [\beta_i(u_j)]_{ij} \in K^{n
  \times n}$ formed by evaluating the~$i$-th boundary condition~$\beta_i$ on
the~$j$-th fundamental solution~$u_j$. So here we assume~$u_1, \dots, u_n$ is a
basis of~$\ker{T}$ and~$\beta_1, \dots, \beta_n$ a basis
of~$\bspc$. Unfortunately, this is not a very intuitive description of~$P$, and
it is not evident how to generalize it to the LPDE case. We have to gain a more
conceptual perspective at~$\beta(u)$ for making the generalization transparent.

Let us write~$\Ev\colon \bspc \oplus \ker{T} \to K$ for the bilinear operation
of evaluation $(\beta, u) \mapsto \beta(u)$. Choosing bases~$\beta_1, \dots,
\beta_n$ for~$\bspc$ and~$u_1, \dots, u_n$ for~$\ker{T}$, the coordinate matrix
of~$\Ev$ is clearly~$\beta(u)$. By the usual technique of dualization, we can
also think of~$\Ev$ as the map~$\bspc\colon \ker{T} \to \bspc^*$ that sends~$u
\in \ker{T}$ to the functional~$\beta \mapsto \beta(u)$. But this map is nothing
else than the \emph{restriction of the trace map}~$\trc\colon \galg \to \bspc'$
to~$\ker{T} \subset \galg$. It is easy to check that the restricted trace is
bijective and that its inverse gives rise to the projector.

\begin{myproposition}
  \label{prop:proj}
  Let~$(T, \bspc)$ be a regular boundary problem with~$E\colon \ker{T} \to
  \bspc'$ being the restricted trace map. Then~$E$ is bijective with the state
  operator~$H$ as its inverse, and $P = H \circ \trc$ is the projector
  with~$\im{P} = \ker{T}$ and~$\ker{P} = \orth{\bspc}$.
\end{myproposition}
\begin{myproof}
  Given any boundary data~$B \in \bspc'$, we know that~$u = H(B)$ satisfies the
  inhomogeneous boundary conditions~$\beta(u) = B(\beta)$ so that~$u^* =
  B$. Hence~$E\colon \ker{T} \to \bspc'$ is a left inverse of~$H\colon \bspc'
  \to \ker{T} \subset \galg$. But it is also a right inverse because~$u \in
  \ker{T}$ is a solution of the semi-homogeneous boundary problem with boundary
  data~$B = u^* \in \bspc'$ and thus must coincide with~$H(B) \in \ker{T}$ by
  the uniqueness of solutions.

  For the projector, observe first that~$\trc\colon \galg \to \bspc'$ is a left
  inverse of~$H\colon \bspc' \to \galg$. As is well
  known~\cite[(A.16)]{RegensburgerRosenkranz2009}, this implies at once that~$P
  = H \circ \trc\colon \galg \to \galg$ is a projector with~$\im{P} = \im{H}$
  and~$\ker{P} = \ker(\trc)$. It remains to show~$\im{H} = \ker{T}$
  and~$\ker(\trc) = \orth{\bspc}$. The second identity follows from the
  definition of~$\trc$. For the first identity, the inclusion from left to right
  follows from the definition of~$H$, the reverse inclusion by writing~$u =
  H(Eu)$ for~$u \in \ker{T}$.
\end{myproof}

We observe that the formula~$P(u) = H(u^*)$ has a very \emph{natural
  interpretation}: The kernel projector picks up the boundary data~$u^*$ of an
arbitrary function~$u \in \galg$ and then constructs the required kernel
element~$H(u^*) \in \ker{T}$ by solving the semi-homogeneous boundary problem
with boundary data~$u^*$. In the LODE case, the relation~$P = H \circ \trc$
reduces to the aforementioned formulae (see~\cite{RegensburgerRosenkranz2009}
after Proposition6.1) after choosing bases~$u_1, \dots, u_n$ for~$\ker{T}$
and~$\beta_1, \dots, \beta_n$ for~$\bspc$. Apart from its conceptual clarity,
the advantage of Proposition\ref{prop:proj} is that it can also be used in the LDPE
case (see after Lemma~\ref{lem:transport-equation}).

\section{The Cauchy Problem for Analytic Functions}
\label{sec:cauchy-analytic}

At this point we switch from the abstract setting of
Sections~\ref{sec:boundary-data} and~\ref{sec:greens-operators} to the concrete
setting of \emph{analytic functions}. Note that we are dealing with
complex-valued functions of real arguments. This means the ground field is~$K =
\CC$, and~$\galg$ is the integro-differential algebra of entire functions
restricted to real arguments.

More precisely, we shall employ the following conventions for easing the burden
of book-keeping: As elements of~$\galg$ we take all holomorphic functions~$\RR^n
\to \CC$ for any~$n \in \NN$, including the constant functions~$u \in \CC$
for~$n = 0$. In other words, $\galg$ is a direct limit of algebras. Moreover, we
have derivations~$\der_n$ and integrals~$A_n$ for all~$n>0$,
namely~$\der_n(u) = \partial u/\partial x_n$ and
\begin{equation*}
  A_n(u) = \int_0^{x_n} u(\dots, \xi, \dots) \, d\xi,
\end{equation*}
where~$\xi$ occurs at the $n$-th position. Clearly, we have then
integro-differential algebras~$(\galg, \der_n, A_n)$ for every~$n > 0$. In
fact, $\galg$ has the structure of a \emph{hierarchical integro-differential
  algebra}. This notion will be made precise at another occasion; for the moment
it suffices to make the following observations. If~$\finset$ is the sublattice
of the powerset~$\pow(\NN^+)$ that consists of finite sets~$\alpha = \{
\alpha_1, \dots, \alpha_k\}$ and the full set~$\NN^+ = \{ 1, 2, 3,
\dots\}$, we define for~$\alpha \in \finset$ the subalgebras
\begin{equation*}
    \galg_\alpha = \{ f \in \galg \mid \der_i f = 0 \text{ for all $i \notin
    \alpha$} \},
\end{equation*}
consisting of the functions depending (at most) on~$x_{\alpha_1}, \dots,
x_{\alpha_k}$. Then~$(\galg_\alpha, \subseteq)$ is a sublattice of~$(\galg,
\subseteq)$ that is isomorphic to the lattice~$(\finset, \subseteq)$. The bottom
element is of course~$\galg_\emptyset = \CC$, the top element~$\galg_{\NN^+} =
\galg$. We write~$\galg_n$ as an abbreviation for~$\galg_{\{1, \dots, n\}}$.

As in the earlier paper~\cite{RosenkranzRegensburgerTecBuchberger2009}, we add
to this algebraic structure all linear \emph{substitution operators}. In
accordance with the above hierarchical structure, we use the
ring~$\mat{\CC}{*}{*}$ of row and column finite matrices with complex
entries.\footnote{The \emph{usage of complex substitutions} in functions of a
  real argument may sound strange at first. But an analytic function on~$\RR^n$
  is of course also analytic on~$\CC^n$ with values in~$\CC$, so there is no
  problem with this view. For example, the substitution~$(1,i)^*$ sends~$f(x) =
  e^x \in \func{}{1}$ to~$f(x+iy) = e^x \, \cos{y} + ie^y \, \sin{y} \in
  \func{}{2}$. Moreover, complex substitutions are indispensible for specifying
  the general solution of elliptic equations like the Laplace equation.}  This
means any~$M \in \mat{\CC}{*}{*}$ can actually be seen as a finite matrix~$M \in
\mat{\CC}{m}{n}$ with~$m$ rows and~$n$ columns, extended by zero rows and
columns. As usual, we identify~$M \in \mat{\CC}{m}{n}$ with the linear
map~$M\colon \CC^n \to \CC^m$, yielding the substitution operator~$M^*\colon
\func{}{m} \to \func{}{n}$ defined by~$u(x) \mapsto u(Mx)$.

We write~$\galg[D,A]$ for the \emph{PIDOS algebra} generated over~$\CC$ by the
operators~$\der_n, A_n \; (n>0)$, the substitutions~$M^*$ induced by~$M \in
\mat{\CC}{*}{*}$ and the exponential basis polynomials~$x^\alpha e^{\lambda x}
\in \galg$. Here~$x$ denotes the arguments~$x = (x_1, \dots, x_n)$ for any~$n
\ge 0$, with exponents~$\alpha = (\alpha_1, \dots, \alpha_n) \in \CC$ and
frequencies~$\lambda = (\lambda_1, \dots, \lambda_n) \in \CC$. Obviously,
$\galg[D,A]$ acts on~$\galg$, with~$D = (D_1, D_2, \dots)$ acting as~$\der_1,
\der_2, \dots$ and~$A = (A_1, A_2, \dots)$ as~$A_1, A_2, \dots$, similar
to the univariate case in the older notation of~\cite{Rosenkranz2005}. Here we
avoid the notation~$\cum$ for the integrals since the powers~$\cum^n$ might be
mistaken as integrals with upper bound~$n$.

The algebra~$\galg[D,A]$ can be described by a \emph{rewrite system} (PIDOS =
partial integro-differential operator system), analogous to the one given
in~\cite{RosenkranzRegensburgerTecBuchberger2009}. We will present this system
in more detail---in particular proofs of termination and confluence---at another
occasion.

Since in this paper we restrict ourselves to the analytic setting, we can appeal
to the well-known \emph{Cauchy-Kovalevskaya
  theorem}~\cite[Thm.~2.22]{RenardyRogers2004} for ensuring the existence and
uniqueness of the solution of the Cauchy problem. While the theorem in its usual
form yields only local results, there is also a global
version~\cite[Thm.~7.4]{Knapp2005} that provides a good foundation for our
current purposes.\footnote{Of course the problem may still be \emph{ill-posed};
  we will not treat this issue here.} Since this form of the theorem is not
widely known, we repeat the statement here.

As usual, we designate one \emph{lead variable}~$t$, writing the other
ones~$x_1, x_2, \dots$ as before. Note that in applications~$t$ is not
necessarily time. The apparently special form of the differential
equation~$Tu=0$ implies no loss of generality: Whenever~$T \in \CC[D]$ is a
differential operator of order~$m$, the change of variables~$\bar{t} = t,
\bar{x}_i = x_i + t$ leads to an equation of the required form.

\begin{theorem}[Global Cauchy-Kovalevskaya]
  \label{thm:global-kovalevskaya}
  Let~$T \in \CC[D_t, D_1, \dots, D_n]$ be a differential operator in
  Caucy-Kovalevskaya form with respect to~$t$, meaning~$T = D_t^m + \tilde{T}$
  with~$\deg(\tilde{T}, t) < m$ and~$\deg(\tilde{T}) \le m$. Then the Cauchy
  problem
  \begin{equation}
    \label{eq:cauchy-problem}
    \left.
    \begin{aligned}
      &Tu = 0\\
      &D_t^{i-1}u(0,x_1, \dots, x_n) = f_{i}(x_1, \dots, x_n) \text{ for $i=1,
        \dots, m$}
    \end{aligned}
    \quad\right\}
  \end{equation}
  has a unique solution~$u \in \func{}{n+1}$ for given~$(f_1, \dots, f_m) \in
  \func{m}{n}$.
\end{theorem}

In the \emph{abstract language} of Sections~\ref{sec:boundary-data}
and~\ref{sec:greens-operators} this is the semi-homogeneous boundary
problem~$(T, \bspc)$ with boundary space
\begin{equation*}
  \bspc = [L_{0,\xi} D_t^i \mid i = 0, \dots, m-1 \;\text{and}\; \xi \in \RR^m],
\end{equation*}
where the evaluation~$u(t, x_1, \dots, x_n) \mapsto u(0, \xi_1, \dots, \xi_n)$
is written as the substitution~$L_{0,\xi} = \diag(0, \xi_1, \dots, \xi_m)^*$
denotes . Hence the solution of~\eqref{eq:cauchy-problem} is given by the state
operator~$(f_1, \dots, f_m) \in \func{m}{n} \mapsto u$ if we identify the
boundary data~$B \in \bspc'$ with its coordinate representation~$(f_1, \dots,
f_m) \in \func{m}{n}$ relative to the above boundary basis~$(L_{0,\xi}
D_t^i)$. In detail, $B\colon \bspc \to \CC$ is the unique linear map
sending~$L_{0, \xi} D_t^i \in \bspc$ to~$f(\xi) \in \CC$; confer
Lemma~\ref{lem:trc-coord} for the uniqueness statement. In the sequel these
identifications will be implicit.

For future reference, we mention also that the usual Taylor polynomial allows
one to provide a natural \emph{interpolator} for the initial data,
namely
\begin{equation}
  \label{eq:fund-intp}
  \fri{\bspc}(f_1, \dots, f_m) = f_1(x) + t \, f_2(x) + \cdots +
  \tfrac{t^{m-1}}{(m-1)!} \, f_m(x),
\end{equation}
which we will not need here because compute the kernel projector directly from
its first-order factors.

In this paper, we will study the Cauchy problem~\eqref{eq:cauchy-problem} for a
\emph{completely reducible operator}~$T(D)$, meaning one whose characteristic
polynomial~$T(\lambda) \in \CC[\lambda] = \CC[\lambda_1, \dots, \lambda_n]$
splits into linear factors. Hence assume~$T = T_1^{m_1} \cdots T_k^{m_k}$ with
first-order operators~$T_1, \dots, T_k \in \CC[D]$. By a well-known consequence
of the Ehrenpreis-Palamodov theorem, the general solution of~$Tu=0$ is the sum
of the general solutions of the factor equations~$T_1^{m_1} u = 0, \dots,
T_k^{m_k} u = 0$; see the Corollary on~\cite[p.~187]{Hansen1983}. Hence it
remains to consider differential operators that are powers of first-order ones
(we may assume all nonconstant coefficients are nonzero since otherwise we
reduce~$n$ after renaming variables).

\begin{mylemma}
  \label{lem:power-first-order}
  Let~$T = a + a_0 \der_t + a_1 \der_1 + \cdots + a_n \der_n \in \CC[D]$ be a
  first-order operator with all~$a_i \ne 0$. Order the variables such that all
  cumulative sums~$a_0 + a_1 + \cdots + a_{i-1}$ are nonzero. Then the general
  solution of~$T^m u=0$ is given by
  \begin{align}
    \label{eq:fundsys-const-coeff}
    &u(t, x_1, \dots, x_n) = \sum_{i=1}^m c_i(\bar{x}_1, \dots, \bar{x}_n) \,
    \tfrac{t^{i-1} e^{-a t/a_0}}{(i-1)!},\\
    \label{eq:fundsys-varchange}
    & \bar{x}_i = t + x_1 + \cdots + x_{i-1} - (a_0 + a_1 + \cdots +
    a_{i-1}) \, x_i/a_i,
  \end{align}
  where~$(f_1, \dots, f_m) \in \func{m}{n-1}$ are arbitrary functions of the
  indicated arguments.
\end{mylemma}
\begin{myproof}
  This can be found in some textbooks on differential
  equations~\cite[p.~139]{Bhamra2010}. Setting up the change of variables given
  by~\eqref{eq:fundsys-varchange} and~$\bar{t} = t$, the differential operator
  is~$T = (a + a_0 \, \der_{\bar{t}})^s$ in the new coordinates (the ordering
  ensures invertibility). Clearly, its fundamental solutions are~$u(\bar{t}) =
  c_i \, \bar{t}^{i-1} e^{-a \bar{t}/a_0} \; (i = 1, \dots, m)$, and the
  ``integration constants''~$c_i$ are arbitrary functions of~$\bar{x}_1, \dots,
  \bar{x}_n$.
\end{myproof}

In principle, one could now combine the general
solutions~\eqref{eq:fundsys-const-coeff} for each factor, substitute them into
the initial conditions of~\eqref{eq:cauchy-problem} and then solve for the~$c_i$
in terms of the prescribed boundary data~$(f_1, \dots, f_m)$. With this choice
of~$c_i$, the general solution will become the state operator for the Cauchy
problem. However, this is a very laborious procedure, and therefore we prefer to
use another route. Since we assume a completely reducible operator, we can
employ the product representation of Proposition~\ref{prop:bp-product}. In that
case, it remains to consider the case of a \emph{single first-order factor}.

\begin{mylemma}
  \label{lem:transport-equation}
  Let~$T = a + a_0 \der_t + a_1 \der_1 + \cdots + a_n \der_n \in \CC[D]$ be a
  first-order operator with all~$a_i \ne 0$. Then the Cauchy problem~$Tu=0$,
  $u(0, x_1, \dots, x_n) = f(x_1, \dots, x_n)$ has the state operator~$H(f) =
  e^{-at/a_0} \, Z^*$. Moreover, the differential operator~$T$ has the right
  inverse
  \begin{align*}
    &\fri{T} = a_0^{-1} \, e^{at/a_0} \, \tilde{Z}^* \, A_t \, e^{-at/a_0} \, Z^*.
  \end{align*}
  Here~$Z \in \mat{\CC}{n+1}{n+1}$ is the
  transformation~\eqref{eq:fundsys-varchange} with~$\bar{t} = t$,
  and~$\tilde{Z}$ is its inverse.
\end{mylemma}
\begin{myproof}
  This follows immediately from Lemma~\ref{lem:power-first-order}. The right
  inverse is computed using Lemma~3 of~\cite{Rosenkranz2005} after transforming
  the LPDE to a LODE.
\end{myproof}

By Proposition\ref{prop:proj}, we can determine the \emph{kernel projector} for
the Cauchy problem of Lemma~\ref{lem:transport-equation} as~$P = H \circ \trc$,
where~$\trc(u) = u(0, x_1, \dots, x_n)$ in this simple case. Having the kernel
projector and the right inverse~$\fri{T}$ in Lemma~\ref{lem:transport-equation},
the \emph{signal operator} is computed by~$G = (1-P) \fri{T}$ as usual. Now we
can tackle the general Cauchy problem~\eqref{eq:cauchy-problem} by a simple
special case of Proposition~\ref{prop:bp-product}.

\begin{myproposition}
  \label{prop:ivp-product}
  Let~$T_1, T_2 \in \CC[D]$ be two first-order operators with nonzero
  coefficients for~$\der_t$. If~$L_{0,\xi}$ is the evaluation defined after
  Theorem.~\ref{thm:global-kovalevskaya}, then we have
  \begin{equation*}
    (T_1, [L_{0, \xi} \mid \xi \in \RR]) \, (T_2, [L_{0, \xi} \mid \xi \in \RR]) 
    = (T_1 T_2, [L_{0, \xi}, L_{0, \xi} D_t \mid \xi \in \RR]
  \end{equation*}
  for the product of the Cauchy problems.
\end{myproposition}
\begin{myproof}
  By the definition of the product of boundary problems, we have to show
  that~$[L_{0, \xi}, L_{0, \xi} D_t] = [L_{0, \xi}, L_{0, \xi} T_2]$. Since each
  of these is defined as a biortho\-gonal, it suffices to prove that the
  system~$u(0,\xi) = u_t(0,\xi) = 0$ has the same solutions as the
  system~$u(0,\xi) = (T_2 u)(0,\xi) = 0$. But the latter is given by
  \begin{equation*}
    (T_2 u)(0,\xi) = a \, u(0,\xi) + a_0 \, u_t(0, \xi) + a_1 \,
    \tfrac{\partial}{\partial x_1}(0, \xi) + \cdots +
    \tfrac{\partial}{\partial x_n}(0, \xi),
  \end{equation*}
  where the first term and the~$x_i$-derivatives vanish since~$u(0,\xi) =
  0$. Using~$a_0 \ne 0$, this implies that the two systems are indeed
  equivalent.
\end{myproof}

This settles the completely reducible case: Using
Proposition~\ref{prop:ivp-product} we can break down the \emph{general Cauchy
  problem}~\eqref{eq:cauchy-problem} into first-order factors with single
initial conditions. For each of these we compute the state and signal operator
via Lemma~\ref{lem:transport-equation}, hence the state and signal operator
of~\eqref{eq:cauchy-problem} by Proposition~\ref{prop:bp-product}.

\section{Conclusion}
\label{sec:conclusion}

As explained in the Introduction, we see the framework developed in this paper
as the first stage of a more ambitious endeavor aimed at boundary problems for
general constant-coefficient (and other) LPDEs. Following the enumeration of the
Introduction, the next steps are as follows:
\begin{enumerate}
\item Stage (1) was presented in this paper, but the \emph{detailed
    implementation} for some of the methods explained here is still ongoing. The
  crucial feature of this stage is that it allows us to stay within the (rather
  narrow) confines of the PIDOS algebra. In particular, no Fourier
  transformations are needed in this case, so the analytic setting is entirely
  sufficient.
\item As we enter Stage (2), it appears to be necessary to employ stronger
  tools. The most popular choice is certainly the \emph{framework of Fourier
    transforms} (and the related Laplace transforms). While this can be
  algebraized in a manner completely analogous to the PIDOS algebra, the issue
  of choosing the right function space becomes more pressing: Clearly one has to
  leave the holomorphic setting for more analysis-flavoured spaces like the
  Schwartz class or functions with compact support. (As of now we stop short of
  using distributions since that would necessitate a more radical departure,
  forcing us to give up rings in favor of modules.)
\item For the treatment of genuine boundary problems in Stage~(3) our plan is to
  use a powerful generalization of the Fourier transformation---the
  \emph{Ehrenpreis-Palamodov integral representation}~\cite{Hansen1983}, also
  applicable to systems of LPDEs.
\end{enumerate}

Much of this is still far away. But the \emph{general algebraic framework} for
boundary problems from Sections~\ref{sec:boundary-data}
and~\ref{sec:greens-operators} is applicable, so the main work ahead of us is to
identify reasonable classes of LPDEs and boundary problems that admit a symbolic
treatment of one sort or another.

\end{document}